\documentclass[twocolumn,prb,showpacs,color,superscriptaddress,epsfig]{revtex4}
\usepackage{graphicx}
\usepackage{amsfonts}
\usepackage{amsmath}
\usepackage{amssymb}
\usepackage{epsfig}
\begin{document}
\title{Spectral density of the Hubbard-model by the continued fraction method}
\author{R. Hayn}
\affiliation{Laboratoire Mat\'eriaux et Micro\'electronique de
Provence, Facult\'e St.\ J\'er\^ome, Case 142, F-13397 Marseille
Cedex 20, France}
\author{P. Lombardo}
\affiliation{Laboratoire Mat\'eriaux et Micro\'electronique de
Provence, Facult\'e St.\ J\'er\^ome, Case 142, F-13397 Marseille
Cedex 20, France}
\author{K. Matho}
\affiliation{Centre de Recherches sur les Tr\`es Basses
Temp\'eratures, Laboratoire associ\'e \`a l'Universit\'e Joseph
Fourier, CNRS, BP 166, F-38042 Grenoble Cedex 9, France}

\date{\today}

\begin{abstract}
We present the continued fraction method (CFM) as a new microscopic
approximation to the spectral density of the Hubbard model in the
correlated metal phase away from half filling. The quantity expanded
as a continued fraction is the single particle Green function.
Leading spectral moments are taken into account through a set of
real expansion coefficients, as known from the projection technique.
The new aspect is to add further stages to the continued fraction,
with complex coefficients, thus defining a terminator function. This
enables us to treat the entire spectral range of the Green function
on equal footing and determine the energy scale of the Fermi liquid
quasiparticles by minimizing the total energy. The solution is free
of phenomenological parameters and remains well defined in the
strong coupling limit, near the doping controlled metal-insulator
transition. Our results for the density of states agree reasonably
with several variants of the dynamical mean field theory. The CFM
requires minimal numerical effort and can be generalized in several
ways that are interesting for applications to real materials.
\end{abstract}


\maketitle

\section{Introduction} \label{Intro}
The Hubbard Hamiltonian \cite{Hubbard} is certainly the most
important model in the field of strongly correlated electrons. The
spectral function for the addition or removal of a single electron
near half filling serves as a paradigm for the excitation spectrum
of highly correlated electrons in the vicinity of a Mott transition.
It was a great success of the dynamical mean field theory
\cite{DMFT} (DMFT) to connect the high- and low-energy parts of the
spectral function in a non-perturbative solution for arbitrary
interaction strength. Especially, it was confirmed that the coherent
low energy excitations in the metallic phase follow the same dynamics as the Kondo
resonance in the Anderson impurity-model, the other generic
Hamiltonian for correlated electrons that is much better understood.
\cite{Hewson}

The analog to the Kondo resonance in the impurity model is a
quasiparticle (QP) band in the lattice model. Both straddle the
chemical potential $\mu$, \i.e. the lowest excitations are gapless.
In the strong coupling limit, the spectral weight $Z$ of the QP band
is small, relative to two sidebands, further removed from $\mu$.
These sidebands are called the Hubbard bands because they are
roughly reminiscent of the Hubbard-I solution. \cite{Hubbard} In the
doping controlled regime, \cite{Imada} one sideband always overlaps
with the QPs, the other represents true high energy
excitations across the correlation gap.

Hubbard-I is an approximation close to the atomic limit,
but nevertheless taking exact spectral moments of the itinerant propagator
up to the second order into account. It can be considered the
ancestor of the projection technique \cite{Fulde} which systematically
incorporates spectral moments of higher order. These approximations have severe
deficiencies in the low energy sector, unless the moment series
can be effectively summed up. In particular, generating a third pole
in the spectral function from the high energy side alone leads to
uncontrolled results.

When the density of states (DOS), as obtained within DMFT, is
resolved with respect to the wavenumber $k$, more details about the
coexistence of this Kondo resonance with atomic like features in
a lattice system are revealed. The lowest excitations are true Fermi
liquid (FL) QP's: (i) The finite DOS at $\epsilon=\mu$ corresponds
to longlived excitations. (ii) These are located in $k$-space on a
Fermi surface (FS) that satisfies Luttinger's theorem.
\cite{Luttinger} (iii) As function of the distance $k-k_{F}$ from
the FS, the excitation energy has a linear, strongly reduced
dispersion. (iv) The damping is quadratic in $k-k_{F}$ but strongly
enhanced, meaning that the linear and quadratic term are of the same
order at a very small energy scale, the coherence energy
$\Delta^{*}$. The two atomic like excitations turn out to be
strongly damped, even when $k$ is on the FS. Their peaks disperse
with $k$ but spectral tails spread over the entire bandwidth.

The DMFT thus unites atomic and itinerant features in a non
perturbative approximation. It is exact only in dimension
$d=\infty$. As a generic scenario, it is expected to hold down to
$d=2$, albeit with the caveat that the DMFT suppresses additional
structure due to bosonic couplings. Earlier approximations at finite
$d$ already yielded QP's\;\cite{Edwards, GA} and established a
connection to the Kondo effect. \cite{Hewson} The high prestige of
the DMFT is due to its ability to produce a selfconsistent,
numerically manageable approximation to the spectral function for
all energies, in particular to the parameter $Z$ that governs the
low energy sector. This has opened a path to realistic modeling of
correlated materials beyond the Hubbard model in such methods as
LDA+DMFT. \cite{Held}

It is nevertheless desirable for several reasons to pursue
alternative methods in parallel. Firstly, a $k$-independent
selfenergy, which is the proper result at $d= \infty$, does not
allow to explain phenomena that depend on the different symmetry
directions in the Brillouin zone, especially in high temperature
superconductors and other low dimensional systems. Cluster
extensions of the DMFT go in the direction of lifting this
restriction, \cite{Maier} but these generalizations are numerically
even more demanding than the DMFT itself.  The precise solution of a
manybody Kondo problem is required at each iteration step towards
selfconsistency. In practice, when designing the "impurity solver",
a trade-off exists between improving the low energy, low temperature
solution and exactly satisfying global sumrules. Such numerical
problems are presently a bottleneck for extensions of the DMFT to
larger clusters or to LDA+DMFT with charge transfer into ligand
bands. A variational aspect was recently found, which may allow to
circumvent some of the numerical problems. \cite{Potthoff1}

In this paper, we present a continued fraction method (CFM) and
implement it for the doping controlled metallic regime near the Mott
transition. Similar to other recent attempts,
\cite{Potthoff,Matho,Avella,Hayn,Kakehashi} we start from the
projection technique, applied to the $k$-resolved single particle
Green function (GF). The notations are introduced in section
\ref{Model}.  In section \ref{HighE}, the connections between the
moment- and continued fraction- (CF) expansion as well as the Pad\'e
approximant (PA) are established. In the PA, qualitatively important
features of the macroscopic system, such as damping, are missing.
They can only be captured by resummation of the CF to infinite
order. The concept of a terminator function (TF), by which an
approximate resummation is achieved, is common to many methods based
on the CF. As a general scheme, we define our CFM  by allowing only
such TF's that preserve the structure of a truncated CF, however
with complex coefficients. Useful recursion relations, that are
properties of PAs, can thus be carried over and the solution for the
GF can be constrained by high as well as low energy sumrules.

Previous solutions obtained with this ansatz \cite{Matho} were
partly phenomenological, because the strong coupling renormalization
$Z$ needed to be inferred from a separate Gutzwiller approximation,
or else was left open for fitting to experiments.
\cite{Matho1,Byczuk} A closed solution is now achieved by minimizing
the total energy in the presence of sumrules, for which the
necessary selfconsistency loops are introduced. The selfconsistent
$Z$ falls below the Gutzwiller value and has a doping dependence
close to that for the exact Kondo scale. \cite{Rasul} This is now a
true microscopic approximation, depending only on the parameters in
the Hamiltonian.

In sections \ref{LowE},  \ref{Numer} and \ref{ImpGap} we have
investigated the TF's that correspond to adding one or two stages
with complex coefficients to the CF. We show how the FS singularity,
the enclosed Luttinger volume in $k$-space and the FL damping can be
modeled rigorously. We assess the quality of our approximations by
comparing the DOS with the DMFT result for two variants of the
impurity solver, namely the numerical renormalization group (NRG)
\cite{NRG} and the non-crossing approximation (NCA). \cite{Pruschke}

The success of the CFM with respect to the Hubbard model allows to
draw some optimistic conclusions about possible generalizations
towards more realistic models, describing correlation effects in a
multiband electronic environment. This will be outlined as part of
the conclusions.

\section{Hamiltonian, Green function and generalities about continued
fractions} \label{Model}

The Hubbard model for a grand canonical ensemble of electrons on a
lattice of $N$ sites ($N\rightarrow\infty$) is written in the usual notations
\begin{equation}
\hat{H}=H-\mu\hat{N}=\sum_{k,\sigma} (E_k -\mu) c_{k \sigma}^{\dagger} c_{k
\sigma} + U \sum_i n_{i \uparrow}n_{i \downarrow} \; .\label{eq1}
\end{equation}
The kinetic energy consists of itinerant Bloch states with energies
$E_k$ and wavenumber $k$, running through one Brillouin zone.
The bandwidth is $2D$ and, when not specified
otherwise, $D$ is used as
unit of both energy and frequency $(\hbar =1)$.
We formulate the method for an arbitrary density of Bloch states. Numerical examples
later on will be calculated for a semi-elliptic density.

The chemical potential $\mu$
is selfconsistently determined to satisfy the condition

\begin{equation}
    n=\langle\hat{N}\rangle/N=2m=\sum_{\sigma}\overline{\langle c_{k\sigma}^{\dagger}c_{k\sigma}\rangle},
    \label{selfconsist}
    \end{equation}
    where the filling factor $n$ $( 0\leq n\leq 2 )$ is part of the
    input. The chemical potential for the $U=0$
limit is designated as $\mu_{0}$. For $U\neq 0$, the right hand side
is calculated with our method. The overline and the bracket
signify Brillouin zone average and ensemble average, respectively.
The filling factor per spin direction in the spin degenerate phase is $m=n/2$.

We approximate the advanced single particle GF
\begin{equation}
    G(k,\omega )=i\int_{-\infty}^{0} e^{i\omega t} \langle
    [c_{k\sigma}(t),c_{k\sigma}^{\dagger}(0)]\rangle,
    \label{GFeq}
\end{equation}
from which the momentum distribution $\langle
c_{k\sigma}^{\dagger}c_{k\sigma}\rangle$ and other observables are
calculated. The spin index is dropped in the unpolarized phase. The
time dependent fermionic destruction operator $c_{k\sigma}(t)$ is in
the Heisenberg representation with $\hat{H}$ and the square bracket
is the anticommutator.  The complex frequency $\omega $ has $\mu$ as
origin. Asymptotically, for large $\omega$, we have $G(k,\omega)
\simeq 1/\omega$. The coefficient 1 reflects the moment $M_{0}=1$ or
spectral norm, as required by the Pauli principle.  For this
relation between the leading coefficient and the norm to remain
valid in an approximation, it is necessary and sufficient to
conserve the Herglotz property. In the case of the advanced GF, it
means that the relation $\mathrm{Im} G(k,\omega )>0$ must be obeyed
throughout the entire halfplane $\mathrm{Im}\omega <0$. The physical
meaning of the Herglotz property is causality and it automatically
entails the existence of Kramers-Kronig relations between the real
and imaginary parts. A great advantage of our method is the
possibility to make straightforward evaluations along the real axis.
Since this limit has to be approached from within the domain of
analyticity, the notation $\omega =\epsilon -i0^+$ with real energy
$\epsilon=E-\mu$ is introduced. The $k$-resolved spectral function
is
\begin{equation}
    A(k,\epsilon)=\frac{1}{\pi} \mathrm{Im} G(k,\epsilon -i0^+). \label{eqA}
\end{equation}
At $U=0$ it has a single sharp peak at the excitation energy
\begin{equation}
    \eta_{k}=E_{k}-\mu_{0}\propto (k-k_{F}),
     \label{eta}
\end{equation}
which also serves to measure distance in $k$-space, at least in the
vivinity of the FS.

The CF expansion, on which our method is based in a crucial way, has
already a long tradition in solid state physics, in the one electron
problem with disorder  \cite{Haydock} as well as in the many
electron problem. \cite{Dagotto} The CF is generated by various
procedures like tridiagonalization, recursion- or Lanczos-methods.
The Hubbard-I GF is the simplest example of a CF that has been
truncated at low order. The exact GF for the Hubbard model on finite
clusters is a CF which naturally ends at very high order.  The CF
for the infinite system does not end. Properties of the
thermodynamic limit, such as damping due to electron-electron
scattering, emerge only after resummation of the CF. Approximate
resummation is achieved by the TF, an analytic function which also
has the Herglotz property.

A well chosen TF is thus expected to bring two improvements to the
approximation for the GF in dimensions $d\geq 2$: (i) From a set of
discrete, more or less intense and more or less densely spaced Dirac
peaks emerges the final shape of the continuous spectral density
(see Ref.~\onlinecite{Treglia} for tight-binding like models and
Ref.~\onlinecite{Kuzian} for strongly correlated electrons). (ii) A
Fermi surface (FS) discontinuity emerges in the momentum
distribution $\langle c_{k\sigma}^{\dagger}c_{k\sigma}\rangle$ at
temperatures below the strong coupling energy scale $\Delta^{*}$.
The FL discontinuity and the correct FS volume will be incorporated
in our ansatz. This means, we take the Luttinger theorem for granted
and use it as a principle, even for strong coupling where there is
no rigorous proof. The energy $\Delta^{*}$ then comes out as part of
the selfconsistent solution.

\section{High-energy part} \label{HighE}

The first moment or center of gravity of $G(k,\omega)$ is
\begin{equation}
    \omega_1 = E_k + m U - \mu .
    \label{eq1c}
    \end{equation}
It disperses like the unrenormalized Bloch energy $E_{k}$. In models
with a more general interaction, a $k-$dependent Hartree-Fock shift
is also present which, for onsite repulsion, reduces to a constant
Hartree shift $mU$. The selfconsistent $\mu$ is the only unknown.

The high energy  expansion about the center of gravity is
\begin{equation}
G(k,\omega)=\frac{1}{\omega-\omega_{1}}+\frac{M_2}{(\omega-\omega_{1})^3}
+\frac{M_3}{(\omega-\omega_{1})^4}+ \ldots.
\label{HighEE}
\end{equation}
Its coefficients
\begin{equation}
M_\lambda = \int_{-\infty}^{\infty} d \epsilon
\cdot (\epsilon - \omega_{1})^\lambda
 \, A(k,\epsilon); \; \lambda =2,3\ldots \label{eq4}
\end{equation}
are called the central moments ($M_{1}=0$, by definition). They can
be related to correlation functions which occur in the short time,
or Liouville expansion of the operator $c_{k\sigma}(t)$ and are
evaluated in the limit $t=0$.

It is remarkable that the
variance $s_{2}$ of $A(k,\epsilon)$, defined by the second central moment
\begin{equation}
    M_{2}=s_{2}^{2}=m(1-m)U^{2}
    \label{variance}
\end{equation}
is $k-$independent
in any dimension $d$, not only $d= \infty$.
All the terms in the high energy
expansion are sensitive to the
low energy sector, be it only via the selfconsistent $\mu$.

We now turn to the CF expansion which is closely related to the moment expansion.
Formally, it is initiated by using $\omega_{1}$ and $s_{2}$ to
write the GF as
\begin{equation}
    G(k,\omega)^{-1}
    = \omega - \omega_{1}-s_{2}^{2}G_{1}(\omega).
    \label{eq2}
    \end{equation}
In this identity, $G_{1}(\omega)$ is again a Herglotz function with
asymptotics  $G_{1}(\omega) \simeq 1/\omega$. Iterations, pushing
the CF further down step by step, require knowledge of the center of
gravity $\omega_{2l-1}$ and the variance $s_{2l}$ of
$G_{l-1}(\omega)$, to write
\begin{equation}
    G_{l-1}(\omega)^{-1} = \omega -
    \omega_{2l-1}-s_{2l}^{2}G_{l}(\omega), l=2,3\ldots.
    \label{eq3}
    \end{equation}
The two new expansion coefficients depend only on the central
moments $M_{\lambda}$ up to the order $\lambda=2l-1$ and $\lambda=2l$
of their respective index.

By truncating the CF, \i.e. by setting
$G_{l}(\omega) \equiv 0$, an approximation to the GF is obtained that has
$l-1$ zeros and $l$ poles on the real axis. This is defined
\cite{Baker} as the PA
$\langle l-1\vert l\rangle $. It represents
the optimal use one can make of a set of known spectral moments up to $M_{2l-1}$.
The present task, constructing the GF, is rendered essentially more
difficult, because the moments themselves are not yet known.
A solution based on a PA can be made selfconsistent
but the moments turn out to be numerically quite inexact.
This fact is often ignored when it is claimed that
a certain high energy approximation obeys a set of "exact" sumrules.

We now discuss some well known results concerning
approximations at the second stage of the CF.  As a still exact representation
of the GF we have
\begin{equation}
 G(k,\omega)=\frac{1}{\displaystyle \omega-\omega_1-\frac{s_2^2}
 {\displaystyle \omega-\omega_3-s_{4}^{2}G_{2}(\omega)}}.
\label{eq3a}
\end{equation}
The relations between the first few terms are:
\begin{eqnarray}
s_2^2 &=& M_2  \nonumber \\
\omega_3 &=& \omega_1 + M_3/M_{2}   \nonumber \\
s_4^2 &=&  M_{4}/M_{2} - M_{2} - (M_{3}/M_{2})^{2}. \label{eq5}
\end{eqnarray}
Besides the variance, quantities used to further characterize the
internal shape of a spectrum are the skewness $\gamma
=M_{3}/M_{2}^{3/2}$ and the kurtosis $\kappa = M_{4}/M_{2}^{2} - 3$.
In terms of these, we have $\omega_{3}=\omega_{1}+\gamma s_{2}$ and
$s_{4}^{2}=s_{2}^{2}(\kappa +2 - \gamma ^{2} )$. From the third
moment one finds the coefficient
\begin{equation}
\omega_3 = (1-m)U+B_3 -\mu \label{eq6}.
\end{equation}
This CF coefficient is the first quantity in the expansion with a
non trivial $k$-dependence. The full correlation function appearing
in the third moment was first derived in  Ref.\
\onlinecite{xxx} and determined
selfconsistently for a short linear chain in Ref.\
\onlinecite{Mehlig}. The shift in the spectral skewness, caused by
$B_{3}$, regulates the dynamical weight
 transfer between the Hubbard peaks at finite $U$. \cite{Meinders}
One can decompose
\begin{equation}
B_3= W_0+W_3(k) \label{eq6a}
\end{equation}
in such a way that the term $W_3(k)$ vanishes in high dimensions.
For making contact with the DMFT we will presently neglect it and
adopt the  expression \cite{xxx}
\begin{equation}
B_3= \frac{(2m-1)}{2m(1-m)} \langle \hat{T} \rangle,
\label{eqB3}
\end{equation}
by which it is linked selfconsistently to the expectation value
of the kinetic energy $\langle \hat{T} \rangle$.

Concerning the behavior of the fourth moment, not even the
correlation functions involved in its selfconsistency loop have as
yet been evaluated. Again, the actual numerical value of $s_{4}^{2}$
is also expected to be sensitive to the low energy sector and, in
low dimensional systems, $k$-dependent.

Given this situation, approximations on the level of Eq.\
(\ref{eq3a}) are at present inevitable. Straightforward truncation,
$G_{2}(\omega)=0$, leads to the PA $\langle 1\vert 2\rangle $. This
solution with two Dirac peaks goes beyond Hubbard-I, because the
dynamical weight transfer is taken into account. The first example
of an approximate resummation of the CF is the alloy analogy,
developed in the paper called Hubbard-III. \cite{Hubbard3} Following
Hubbard's notation, we approximate $s_{4}^{2}G_{2}(\omega)$ by a
$k$-independent TF $\Omega_H(\omega)$, which has to be a Herglotz
function.

The alloy analogy satisfies at least the task (i) of a TF, namely to
generate finite damping. Far away from $\mu$, where the excitations
are incoherent, it actually represents a physically correct picture.
We therefore keep the result $\Omega_H(\omega) \rightarrow iD$ for
large $\omega$ from Hubbard-III. The physical reason, why the
damping is of the order of the bare bandwidth $D$ is that the mean
free path is as short as one lattice constant. In practice, we
incorporate the high energy damping in an effective $\omega_{3}$
\begin{equation}
\bar{\omega}_3 = \omega_3 +iD
\end{equation}
and henceforth deal with a terminator that decays as $1/\omega$.
This way, we conserve the sumrules, encapsuled in the central
moments $M_{0}$ to $M_{3}$. Since Hubbard-III is unrealistic at low
energies, we do not pursue it any further. Nevertheless, it should
be noted that Hubbard-III generates a branchcut in
$\Omega_H(\omega)$, causing the imaginary part to drop back to zero
and a correlation gap with sharp edges to appear, at least in the
zero temperature limit. This property of Hubbard-III is also not
expected to survive in improved approximations for the metallic
phase. We will address the consequences that the absence of a
branchcut has for the shape of the DOS, both in the CFM and in the
DMFT.

To sum up, our approximation to the GF is formally similar to Hubbard-III,
\begin{equation}
G(k,\omega)=\frac{1}{\displaystyle
\omega-\omega_1-\frac{s_2^2}{\displaystyle
\omega-\bar{\omega}_3-\Omega(\omega)}}, \label{eq3b}
\end{equation}
but with a TF, $\Omega_H(\omega)=\Omega(\omega)+iD$, that retains
the strong damping of the alloy analogy only at high energy. Two
successive implementations of the TF with appropriate FL properties
at low energy, are the subject of the following sections.

\section{Low-energy part}\label{LowE}

A FS discontinuity is strictly realized only in the zero temperature
limit and in a system with no residual disorder. Since $T=0$
solutions are hardest to obtain with DMFT  and, on the contrary,
easily implemented with our method, we concentrate in the following
on this limit. We write the standard microscopic definition of a
selfenergy as a complex correction to the bare excitation
$\eta_{k}$:
\begin{equation}
G(k,\omega)^{-1} = \omega-\eta_k-\Sigma(k,\omega) \;
\label{eq9}
\end{equation}
and compare with the inverse of Eq.\ (\ref{eq3b}). The high energy
limit $\Sigma(k,\infty)$ is the difference between two dispersive
quantities. In the present case, Eqs.\ (\ref{eta}) and (\ref{eq1c})
have identical dispersion and
\begin{equation}
p_1=\eta_k-\omega_1=\mu - \mu_{0} -mU \label{eqp1}
\end{equation}
is, in fact, constant.
Within the other approximations, discussed in the preceding section,
we then obtain the $k$-independent selfenergy
\begin{equation}
\Sigma(\omega)=-p_{1}+\frac{s_2^2}{\omega-\bar{\omega}_3-\Omega(\omega)}.
\label{eqSE}
\end{equation}
In this case, as in the DMFT, the  FS has the exact shape of the
uncorrelated system. It is given by all $k-$points where
$\eta_{k}=0$ in Eq.~(\ref{eta}). The QP peak of weight $Z$ at the
Fermi level and the step of amplitude $Z$ in the momentum
distribution are fixed by the conditions
\begin{equation}
    \Sigma(0)=0
    \label{FLcond1}
\end{equation}
and
\begin{equation}
   \frac{ d \Sigma}{d \omega} (0)=\alpha = 1-1/Z <0.
    \label{FLcond2}
\end{equation}
At finite $T$ or in the presence of a residual diffusive mean free
path, $\mathrm{Im} \Sigma (0)$ remains finite.

Guided by the insight that the strong coupling peak is distinct from
the Hubbard peaks, we can formulate a minimal ansatz for the TF
\cite{Matho} as
\begin{equation}
\Omega(\omega)=\frac{(\bar{s}_4)^{2}}{\omega-\bar{\omega}_5}.
\label{TF1}
\end{equation}
Adding a new stage to the CF is the proper way to "add" a pole to
the GF. When this TF is inserted in Eq.\ (\ref{eq3b}), it generates
a GF with three complex zeros in the denominator and two zeros in
the nominator, \i.e. the same structure as the PA $\langle 2\vert
3\rangle $. The connection of the parameters
 $\bar{s}_4^{2}$ and $\bar{\omega}_5$  to central moments
$M_{4}$ and $M_{5}$ is lost.  In fact, the very existence of moments
beyond $M_{3}$ has been sacrificed by admitting $\bar{\omega}_{3}$,
$\bar{s}_{4}$, and $\bar{\omega}_{5}$ as complex quantities. They
now have to be determined from conditions (\ref{FLcond1}) and
(\ref{FLcond2}).

For the Herglotz property one finds
\begin{equation}
(\mathrm{Im} \bar{s}_{4})^{2}/\mathrm{Im}\bar{\omega}_{5}\leq
\mathrm{Im}\bar{\omega}_{3}=  D \label{Herglotz1}
\end{equation}
as a necessary and sufficient condition. This causes all three poles
to lie in the upper half-plane. Further, it guarantees a normalized,
positive semidefinite $A(k,\epsilon)$, which also implies quite
intricate relations between the complex residues.

Now, the important point is the following: This simple ansatz
is so heavily constrained by sumrules that it offers a
selfconsistent solution of the problem, without any free parameters.
It remains to substantiate this claim and then to discuss the quality
of the solution.

After inserting Eq.\ (\ref{TF1}) in Eq.\ (\ref{eqSE}), the
conditions (\ref{FLcond1}) and (\ref{FLcond2}) can be brought into a
system of two linear equations for the unknowns $\bar{s}_4^{2}$ and
$\bar{\omega}_5$. The determinant of this system is
\begin{equation}
det_2=-p_1^2-\alpha s_2^2 \; , \label{eqdet2}
\end{equation}
and the Herglotz property requires $det_{2} > 0$. This is a
constraint on the QP weight $Z$: In stead of $Z\leq 1$ (Pauli
principle) we have $ Z< s_{2}^{2}/(s_{2}^{2}+p_{1}^{2})$. Closer
inspection reveals that it means the QP cannot take more spectral
weight than the peak in the PA $\langle 1\vert 2\rangle $ that is
nearest to $\mu$. Since around half filling this weight stays above
$1/2$, it is indeed only a weak constraint.

The solution
\begin{equation}
\bar{s}_4^2=\frac{p_2^2}{det_2} \; \label{eqs4},
\end{equation}
and
\begin{equation}
    \bar{\omega}_5=-\frac{p_1 p_2}{det_2}  \label{eqom5}
\end{equation}
is expressed in terms of the complex quantity
\begin{equation}
p_2=-(p_1 \bar{\omega}_3 + s_2^2 ). \label{eq17}
\end{equation}
It fulfills the Herglotz condition (\ref{Herglotz1}) with the
equality sign. This is a consequence of our strong $T=0$ constraint
$\Sigma(0)=0$, concerning both the real and imaginary part. The
selfenergy is now parametrized, up to $Z$, which remains free within
a restrained interval and will be determined by minimizing the total
energy.

We note, before closing this section, that Ref. \onlinecite{Matho}
allows to define one-pole TF's for the more general case of a
truncated GF that is expressed as a higher order PA. The general
algorithm is given, by which the Eqs.\ (\ref{FLcond1}) and
(\ref{FLcond2}) can be fulfilled.

\section{Numerical procedure} \label{Numer}

The uncorrelated chemical potential as function of the filling,
$\mu_{0}(n)$, depends only on the kinetic energy part and is
determined once for all.
The DOS per lattice site in the $U=0$ limit
\begin{equation}
\rho_0(\varepsilon)=\frac{2}{\pi}\mathrm{Im}
F_{0}(\varepsilon-i0^{+})
\end{equation}
is obtained from the onsite GF
\begin{equation}
    F_{0}(\omega)=\frac{1}{N} \sum_k \frac{1}{\omega+\mu_{0}-E_{k}}.
    \label{eq1b}
\end{equation}
A factor two comes from summing over spin directions. The DOS of the
correlated system
\begin{equation}
\rho(\varepsilon)=\frac{2}{\pi}\mathrm{Im} F(\varepsilon-i0^{+})
\end{equation}
is obtained from $F(\omega)=\overline{ G(k,\omega)}$, the on-site GF
in real space, which is independent of the site index. For a
$k$-independent selfenergy such as Eq.\ (\ref{eqSE}), the on-site
GF's $F(\omega)$ and $F_{0}(\omega)$ are related to each other by
\begin{equation}
F(\omega)=F_0(\omega-\Sigma(\omega)).
\end{equation}
The $k$-summations can then be
carried out by using the analytic function that represents the
solution for $F_{0}(\omega)$ in the limit $N\rightarrow\infty$.

We now turn to the discussion of the selfconsistency loops.
 The condition for $\mu$ is
implemented at $T=0$ by the integral
\begin{equation}
 \int_{-\infty}^{0} d \epsilon \rho(\epsilon) = n.
 \label{sceq1}
\end{equation}

According to Eq.\ (\ref{eqB3}), the term $B_3$ from the third moment
requires the selfconsistent determination of the kinetic energy,
$\langle \hat T \rangle = 2 \int_{-\infty}^{0} d \epsilon \overline{
A(k,\epsilon) E_k}$. One finds
\begin{eqnarray}
\langle \hat T \rangle &=&  \frac{2}{\pi} \int_{-\infty}^{0}d
\epsilon \ \mathrm{Im} (\tilde{\omega}F_0(\omega-\Sigma(\omega))-1)
\\
\tilde{\omega}&=&\epsilon-i0^{+} +\mu_0-\Sigma(\epsilon-i0^{+}). \nonumber
\label{sceq2}
\end{eqnarray}
Finally, the total energy is
\begin{equation}
E_{tot}=\frac{1}{2}(\langle \hat T \rangle + \int_{-\infty}^{0}d
\epsilon\;\epsilon \;\rho(\epsilon)).
\label{eqEtot}
\end{equation}
The integrals in (\ref{sceq1}) - (\ref{eqEtot}) are carried out
numerically. For the calculations in this paper we took the on-site
GF

% Caption Konrad
%\begin{figure}
%\includegraphics[width=6.5cm,angle=-90]{fig1.eps}
%\caption{Energy minimum $E_{tot}(Z)$ for $U=4$, $D=1$,
%$\mu_0=-0.166$ corresponding to $n=0.79$. } \label{fig1}
%\end{figure}

% Caption Roland
\begin{figure}
\includegraphics[scale=0.35,angle=-90]{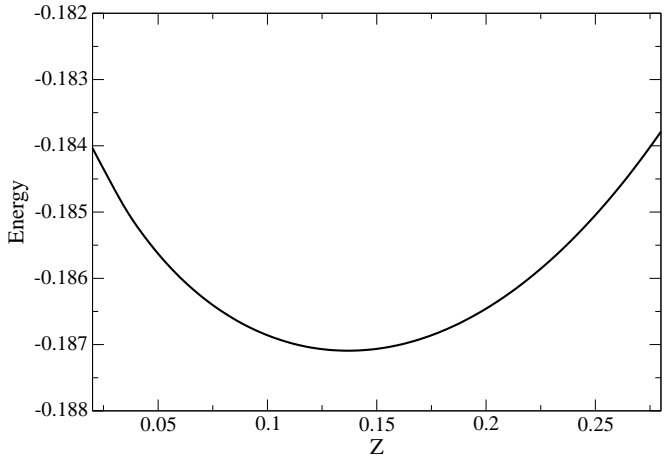}
\caption{Energy minimum $E_{tot}(Z)$ for $U=4$, $D=1$, and
$\mu_0=-0.166$ corresponding to $n=0.79$. The choice of $D=1$ means
that all energies are measured in units of the halfwidth $D$ of the
uncorrelated band.} \label{fig1}
\end{figure}

\begin{equation}
F_0(\omega)=\frac{2}{(\omega+\mu_{0})
\left(1+\sqrt{1-\frac{1}{(\omega+\mu_{0})^2}}\right)}.
\end{equation}
In the context of $d=\infty$, it is the GF for a Bethe lattice. A
halfwidth $D=1$ is now used as energy unit. For the Herglotz
property, it is important to choose the square root with a positive
real part. The model DOS belonging to this GF,
\begin{equation}
\rho_0(\varepsilon)=\frac{4}{\pi}\sqrt{1-(\epsilon+\mu_{0})^2}
\end{equation}
is the semi-elliptic function which was also used by Hubbard.

While searching for the selfconsistent $\mu$ and $\langle \hat T
\rangle$ at a given input $n$ and $U$, the renormalization  $Z$ is
still kept as a parameter, only limited by the condition
$det_{2}>0$. With these constrained solutions for the GF, we
calculate the total energy $E_{tot}$. As shown in the example of
Fig.\ \ref{fig1}, $E_{tot}$ has a well defined minimum as a function
of $Z$. Taking the value which minimizes $E_{tot}$ fixes the last
parameter $Z$ and defines our solution for the GF.

% Caption Konrad
%\begin{figure}[ht]
%\includegraphics[width=7.5cm,angle=-90]{fig2.eps}
%\caption{Spectral density, comparison of $\rho_0(\epsilon)$ and
%$\rho(\epsilon)$. } \label{fig2}
%\end{figure}

% Caption Roland
\begin{figure}
\includegraphics[scale=0.35,angle=-90]{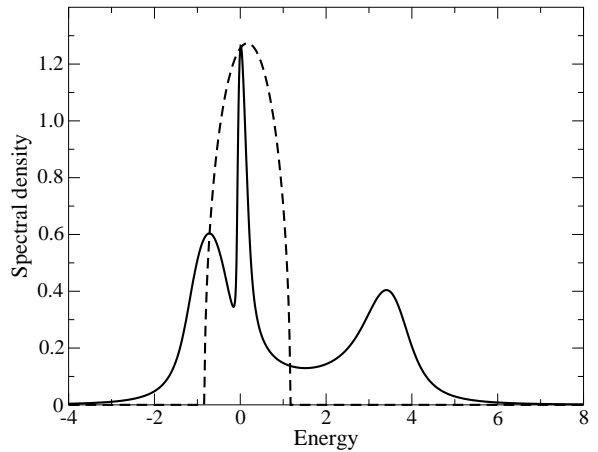}
\caption{Spectral density, comparison of $\rho_0(\epsilon)$ (dashed
line) and $\rho(\epsilon)$ (full line) for the same parameters as in
Fig.\ \ref{fig1}. } \label{fig2}
\end{figure}

The DOS obtained for the same input as in Fig.\ \ref{fig1} is shown
in Fig.\ \ref{fig2}, together with the $U=0$ limit. The QP band has
the same intensity at the Fermi level as the uncorrelated band,
$\rho(0)=\rho_0(0)$. This invariance signals the unitary limit for
the Kondo resonance in the limit $T=0$. Thus, the reduction of the
QP weight does not show up in $\rho(0)$ but in the bandwidth, which
is scaled down by $Z$. In a lattice system, this one-to-one
relationship between QP weight and bandwidth only holds when the
selfenergy is local ($k$-independent).

On the $k$-resolved level, near the FS, the QP-pole in the complex
plane has a parabolic trajectory parametrized by $\eta_{k}$, Eq.\
(\ref{eta}):
\begin{equation}
    \omega^{*}(k)=Z\eta_{k}+i\Gamma_{k},
\end{equation}
with a scattering rate
\begin{equation}
    \Gamma_{k}= (Z\eta_{k})^{2}/\Delta^{*}.
    \label{Gamma}
\end{equation}
The halfwidth for coherent states within the QP band is
\begin{equation}
\Delta^{*}=\frac{Zs_{2}^{2}\vert p_{2}\vert
^{2}}{D((1-Z)s_{2}^{2}-Zp_{1}^{2})^{2}}.
\label{eq*}
\end{equation}
This formula is well behaved also in the weak coupling limit, in
fact for all possible metallic, unpolarized regimes of the Hubbard
model. In the strongly correlated regime, the energy scale
$\Delta^{*}$ is smaller than $ZD$, so that the excitations in the
wings of the QP band cease to be coherent. Since we have modeled the
ballistic limit (residual diffusive scattering rate $\Gamma_{d}=0$)
the QP resonance in $A(k,\epsilon)$ is a Dirac peak for $k=k_{F}$
and, for $k\neq k_{F}$, it has the so called Breit-Wigner
lineshape\; (see Hedin and Lundquist \cite{Hedin} for a generic
plot). This shape is due to an interference between the QP residue
and the other residues. The lineshape becomes approximately
Lorentzian whenever a $\Gamma_{d}>\Gamma_{k}$ is present.

Returning to the DOS, we note that the global shape of the valence
spectrum for a hole doped Mott insulator, \i.e. QP band and lower
Hubbard band, is well rendered by our present approximation. The
sumrules up to $M_{3}$ are exactly satisfied and their interplay
regulates the overall skewness and the relative weight of all three
features.

The one-pole TF has the drawback of being unable to reproduce a sharp gap formation.
The high level of intensity between the
QP band and the upper Hubbard band  shows that
 the dynamical spectral transfer \cite{Meinders} is not realized
 completely, at least for $U/D=4$. The intensity at the minimum
 decays like $(U/D)^{-2}$, so that this spurious effect
disappears for larger $U$. We shall discuss the presence of
residual intensity in the gap region in more detail when we
compare with our second ansatz and with the DMFT.

Two remarks to conclude this section: (i) The limit $U=\infty$
describes spin- and charge-excitations in the subspace of singly
occupied sites. It is equivalent to the $t-J$-model with $J=0$. The
one-pole TF thus allows to project out a quantitatively valid GF for
the valence sector near this limit, up to terms of order
$(U/D)^{-2}$.  (ii)~Ratios $U/D\approx 4$ are relevant for the
doping controlled Mott transition in real materials close to
criticality. Our main motivation to pursue the CFM was to
investigate whether by simply adding a second complex stage to the
TF we could handle this regime in a semiquantitative way. The
derivation of the two-pole TF and its application to $U/D = 4$ are
presented in the next section.

\section{Improving the dynamical weight transfer} \label{ImpGap}

To generalize our ansatz, we introduce algebraic expressions for
$\Omega(\omega)$, such that Eq.\ (\ref{eq3b}) can be cast into the
form of a truncated CF with complex coefficients. This defines the
CFM, provided the Herglotz condition is satisfied. The $k$-resolved
GF has then the structure of a generalized higher order PA. By
terminating the PA $\langle 1\vert 2\rangle $, we still retain the
important sumrules that govern the dynamical weight transfer.
Spectral moments beyond $M_{3}$ cannot be recovered, but this  may
not be a great sacrifice, given the difficulties known from the
projection method to obtain correct values for higher moments.

What can be gained by using complex coefficients is the possibility
to model constructive and destructive interference phenomena in the
GF at intermediate energies. A single feature in the spectral
function can be built up by the contributions of several poles,
resulting in uncommon lineshapes. An ansatz frequently employed in
the phenomenological interpretation of spectra is the superposition
of complex poles with real residues (superposition of Lorentzians in
the spectrum). Although this allows several peaks to coalesce, it
still eliminates interference. One striking example of a Fano like
interference within the coherence range of halfwidth $\Delta^{*}$ is
the Breit-Wigner lineshape of the QP. \cite{Hedin} As discussed in
the preceding section, the complex residues of the two valence
poles in the GF (arising from the one-pole TF in the large $U$ limit)
are enough to obtain this lineshape.

Likewise, the dynamical weight transfer and the formation of the
correlation gap can  be interpreted as a destructive interference in
the intermediate energy range between the Hubbard bands. When the
interference is complete the function $G_{2}(\omega)$ in Eq.\
(\ref{eq3a}) should acquire a branchcut and a gap interval with zero
DOS and sharp edges should result. This may be possible only on the
insulating side of the Mott transition and strictly at $T=0$. When
the system is metallic and the chemical potential falls in a region
of high DOS, it is satisfactory to model the correlation gap by a
deep minimum. We demonstrate here that this situation is captured by
a two-pole TF of the form
\begin{equation}
\Omega(\omega)=\frac{\bar{s}_4^2} {\displaystyle
\omega-\bar{\omega}_5-\frac{\bar{s}_6^2}{\displaystyle
\omega-\bar{\omega}_7}} \; \; . \label{eqTF2}
\end{equation}
The new degrees of freedom are given by $\bar{s}_4^2$ and
$\bar{\omega}_5$. These will be found due to some qualitative
arguments, restricting the ansatz from the start. Then,
$\bar{s}_6^2$ and $\bar{\omega}_7$ can again be eliminated  by the
FL conditions of Eqs.\ (\ref{FLcond1}) and (\ref{FLcond2}), using
the next iteration of the algorithm in Ref.\ \onlinecite{Matho}.

The GF now has four poles and the Herglotz condition becomes a
crucially important issue. To formulate it, for arbitrary complex
values of $\bar{\omega}_3$ to $\bar{\omega}_7$, seems at first sight
rather difficult. The GF on the FS  (Eq.\ (\ref{eq9}) with
$\eta_{k}=0$) has additive coherent and incoherent contributions,
\begin{equation}
\frac{1}{\omega-\Sigma(\omega)}=\frac{Z}{\omega} + G_b(\omega).
\label{eqGb1}
\end{equation}
The decomposition is possible, because one pole lies exactly on the
real axis. This will enable us to manage the Herglotz condition for
$\Sigma(\omega)$ more easily: from Eq.\ (\ref{eqTF2}) we obtain  a
background function $G_b(\omega)$ with three poles that can be
written
\begin{equation}
G_b(\omega)=\frac{1-Z}{\displaystyle
\omega-\Omega_1-\frac{\Sigma_2^2} {\displaystyle
\omega-\Omega_3-\frac{\Sigma_4^2}{\displaystyle \omega-\Omega_5}}}.
\label{eqGb2}
\end{equation}
The new coefficients are designated by capital Greek letters.
Systematically, they depend on $Z$ and, at order $\lambda$, on all
coefficients in Eqs.\ (\ref{eq3b}) and (\ref{eqTF2}) with index
$\lambda^{'}\leq \lambda$. Explicitly, the first three are
\begin{eqnarray}
\Omega_1 &=& - \frac{p_1}{1-Z} \nonumber \\
\Sigma_2^2 &=& \frac{Z \; det_2}{(1-Z)^2}  \\
\Omega_3 &=& \frac{1}{p_1} \left\{ \frac{\alpha \; p_2 \;
s_2^2}{det_2} + \frac{det_2}{\alpha} \right\}, \nonumber
\label{eqGb3}
\end{eqnarray}
in terms of the previously defined quantities, Eqs.\
(\ref{variance}), (\ref{eqp1}), (\ref{eqdet2}), and (\ref{eq17}).

The high energy damping in the background function is
\begin{equation}
\mathrm{Im} \Omega_3 = (1+ \frac{p_{1}^{2}}{det_2}) \mathrm{Im}
\bar{\omega}_3> \mathrm{Im}\bar{\omega}_3=D. \label{eqGb4}
\end{equation}
Between the coherence energy $\Delta^{*}$ in  Eq.\ (\ref{Gamma}) and
the background function at the Fermi edge there is the relation
\begin{equation}
\Delta^{*}\mathrm{Im}G_{b}(0)=Z. \label{eq**}
\end{equation}
For $\Sigma_{4}=\Omega_{5}=0$, we recover the one-pole TF and
Eq.~(\ref{eq*}) for $\Delta^{*}$. Since $\Omega_1$ and $\Sigma_2^2$
are real, there are now only three complex quantities and the
Herglotz condition can be specified exhaustively, analogous to
Eq.~(\ref{Herglotz1}):
\begin{equation}
(\mathrm{Im} \Sigma_{4})^{2}/\mathrm{Im}\Omega_{5}\leq
\mathrm{Im}\Omega_{3}. \label{Herglotz2}
\end{equation}
The foregoing analysis suggests that $\Sigma_{4}$ and $\Omega_{5}$
are more useful than $\bar{s}_{4}$ and $\bar{\omega}_{5}$ as control
parameters. To obtain the selfenergy, one can then use
Eqs.~(\ref{eqGb1}) and (\ref{eqGb2}). For further discussion we
parametrize
\begin{eqnarray}
\Omega_3 &=& X_3 + i Y_3 \nonumber \\
\Sigma_4 &=& X_4 + i Y_4  \\
\Omega_5 &=& X_5 + i Y_5. \nonumber
\end{eqnarray}
A minimal requirement for properly defining the dynamical weight
transfer is vanishing $\mathrm{Im}\Sigma(\epsilon-i0^{+})$ in one
other point  $\epsilon=x_{0}$ on the real axis, apart from
$\epsilon=0$. It happens if (and only if) the equality sign applies
in (\ref{Herglotz2}).

This leads to the condition $Y_4^2=Y_3 Y_5$ and to
\begin{equation}
    x_{0}=X_{5}-\frac{Y_{5}}{Y_{4}}X_{4}
\label{eqx0}
\end{equation}
for the position. The Herglotz property guarantees that it is in
fact a minimum. The influence of this interference on the shape of
the valence spectrum is weakest for $X_5=0$. For simplicity, we also
need to set $Y_5=Y_4=Y_3$, where $Y_3$ is already defined in
Eq.~(\ref{eqGb4}). The last parameter $X_{4}=-x_{0}$ is then fixed
by the point with lowest intensity inside the correlation gap.

% Caption Konrad
\begin{figure}[ht]
\includegraphics[width=7cm,angle=-90]{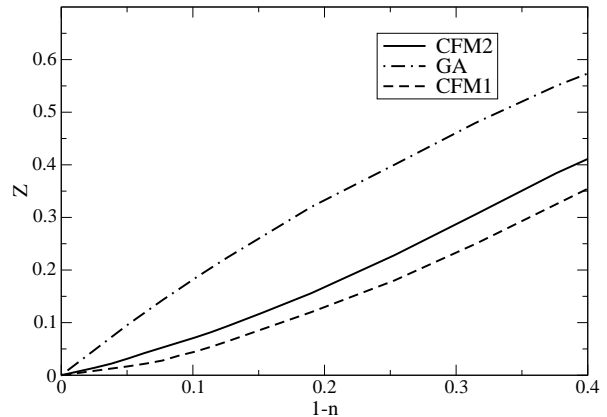}
\caption{Spectral weight $Z$ of QP pole as function of electron
density $n$ for $U=4$. Comparison of the Gutzwiller approximation
(GA) to our result with the one- and two-pole TF's (CFM1 and CFM2).
} \label{fig3}
\end{figure}

Before continuing with this ansatz, it is important to realize that
it cannot apply exactly at half filling. There, the metallic phase
is obtained by driving $U/D$ below the critical ratio (so called
bandwidth controlled transition). \cite{Imada} The particle-hole
symmetric DOS has a quite different morphology than what is shown in
Fig.\ \ref{fig2}: the QP's are in the center and the correlation gap
is split in two symmetric gaps of order $U/2$. \cite{DMFT} In our
approach, it can be envisaged to use one additional complex stage to
model two symmetric destructive interferences.

In the doping controlled regime, there is only one large correlation
gap and we have a good qualitative argument for $x_{0}$: the strong
skewness (large $\vert p_{1} \vert$) causes the QP band and the
minimum position $x_{0}$ to always be on opposite sides of the
center of gravity. This is well satisfied by setting
\begin{equation}
x_0= \mbox{Re} \bar{\omega}_3.
\label{eqzero}
\end{equation}
The remaining free parameter $Z$ is determined again by minimizing
the total energy. The numerical procedure is as described before. In
Fig.\ \ref{fig3}, results with the one- and two-pole TF's (CFM1 and
CFM2) are compared to the Gutzwiller approximation (GA) at constant
$U$, as function of the filling. The upper curve is the well known
lower bound for the GA, $Z=(1-n)/(1-m)$, obtained by excluding
double occupancy. By projecting out the background, the GA is known
to systematically overestimate the coherent weight. The behavior
that results from the CFM, \i.e.  :  lowering of $Z$ and upward
curvature at the approach of zero doping ($1-n\rightarrow 0$), is
close to that of the exact Kondo scale in the Bethe ansatz solution
for the Anderson impurity. \cite{Rasul}

\section{Comparison with other methods}

With the two-pole TF, realistic results for the DOS in the doping
controlled regime can be obtained,  even close to the critical $U$. To
illustrate this, we compare our CFM with the DMFT for two different impurity solvers.
The impurity solvers perform the crucial step in mapping
the Hubbard lattice model onto an Anderson
impurity model. The effective medium surrounding a given site is
determined self-consistently, still a formidable manybody problem.
The NRG, \cite{NRG} used to
solve it at the lowest temperatures and energies, requires a heavy amount of
computer time. The NCA
\cite{Pruschke,Grewe} is an alternative, more analytic method,
less reliable for $\vert \omega \vert \ll \Delta^{*}$, but obeying
high energy sumrules well. Therefore, NRG and NCA are expected to be
complementary.

\begin{figure} [ht]
\includegraphics[scale=0.35,angle=-90]{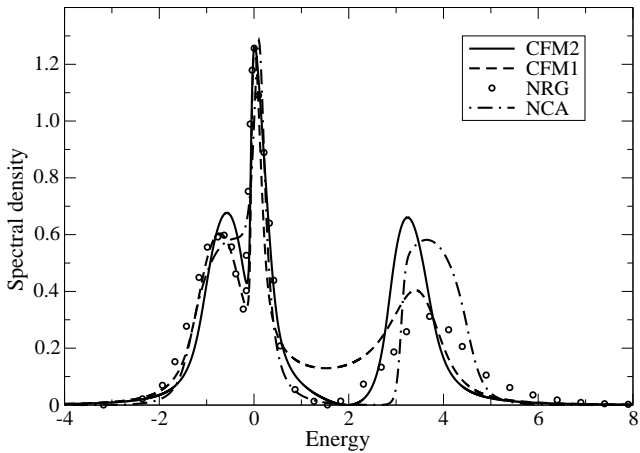}
\caption{Spectral density. Comparison of the continued fraction
method using the one- and two pole TF's (CFM1 and CFM2) with NCA and
NRG (data taken from Ref.\ \onlinecite{Byczuk}). } \label{fig4}
\end{figure}

A comparison for the same parameters as before, i.e.\ $U=4$ and
$n=0.79$, is shown in Fig.\  \ref{fig4}.  The NRG data are taken
from Ref.\ \onlinecite{Byczuk}, NCA is our own unpublished
calculation, CFM1 is again the DOS from Fig.\  \ref{fig2} and CFM2
the result with Eq.~(\ref{eqTF2}). All four solutions obey the
$\rho(0)=\rho_0(0)$ condition. This confirms that temperatures in
the DMFT solutions are sufficiently low to warrant a comparison with
our $T=0$ results. As manifest in the width of the QP band, the
selfconsistent $Z$ obtained for CFM2 coincides with both versions of
DMFT. Since NRG is expected to determine essentially the exact low
energy scale, this is a good point for both the NCA and the CFM2
results.

% Caption Konrad
%\begin{figure} [ht]
%\includegraphics[width=6.5cm,angle=-90]{fig4.eps}
%\caption{Spectral density. Comparison of continued fraction method
%(CFM1 and CFM2) with our own NCA data and NRG
%data taken from Ref.\cite{Byczuk}. }
%\label{fig4}
%\end{figure}

% Caption Roland

The solutions start to differ somewhat in the gap region.
Neither DMFT version shows a gap with
sharp edges that would correspond to a branchcut in the selfenergy.
A real benchmark for low $T$ impurity solvers in the doping
controlled regime does not yet exist. From the NCA, we can confirm that
some very low residual density inside the gap seems to be the generic situation.

In the ansatz for CFM2, the existence of a point with zero DOS is
postulated. Determining its position according to Eq.~(\ref{eqzero})
involves the selfconsistency conditions for $\mu$ and $B_{3}$. The
quantitative agreement with the  DMFT in the QP band and good
overall agreement in the entire valence sector is due to this built
in interference. In comparing CFM1 and CFM2, one notices a feedback
of the improved gap region on the QP band: The sumrules up to
$M_{3}$ are satisfied for both approximations, but the dynamical
weight transfer is more complete within CFM2. Removing the spurious
intensity inside the gap slightly raises the QP weight (Compare
Fig.\ \ref{fig3}), bringing it in agreement with the NRG.

The rather large variation among the different solutions in the
region of the upper Hubbard band is remarkable and still deserves
more detailed investigations.  At higher temperatures, $T\geq
\Delta^{*}$, where Quantum Monte Carlo is available as benchmark,
the NCA was found to be satisfactory. \cite{Pruschke} In the present
comparison, the NCA  comes closer to obeying the sumrules than the
NRG. As far as numerical effort is concerned, the NRG is the most
demanding, followed by the NCA. The CFM2 stands up quite honorably
in this comparison, especially when considering that the sumrules
are rigorously incorporated, no "technical" broadening needs to be
introduced and the required computer time to achieve selfconsistency
is in fact negligible.

\section{Discussion and outlook}

We present the CFM as a new method to calculate the selfenergy, as
well as various $k-$resolved and (partially or fully) $k-$integrated spectral
functions of the Hubbard model in the correlated metal phase. We
expand the single particle Green function as a continued fraction,
as far as moment sumrules are exploitable, and then use a
properly chosen terminator function. In this paper, moment sumrules
up to $M_{3}$ are implemented and the "terminator" is a
$k-$independent complex function
with one or two poles that obeys the correct Fermi liquid properties at
low energies. In this local approximation to the selfenergy,
we compare our results for the density of states with the DMFT. Our method has
a precision comparable to state-of-the-art impurity solvers NRG and
NCA. It covers all energy scales reliably, whereas the low T impurity
solvers each have their strengths and weaknesses.

With the second stage in the terminating
function we are able to improve
the dynamical weight transfer between the upper and lower
Hubbard peak and thereby obtain very good agreement with DMFT for
the QP weight $Z$ or low energy scale.
This is significative, because NRG-DMFT yields the exact result for this quantity.
Unlike the time consuming DMFT calculations, the CFM uses simple, algebraic
functions, for which selfconsistency conditions are rapidly found.

The CFM is generalizable in many directions. However, the
possibility to circumvent heavy manybody calculations by such a
simplified ansatz seems too attractive to be true. Thus, before
advocating possible extensions, we need to analyze the reason for
the quantitative success of the CFM in the strong coupling limit.
The Hubbard model with a local selfenergy is, admittedly, only a toy
model but nevertheless an obligatory testing ground for this
important issue.

The algebraic terminator functions were already introduced earlier.
Their Fermi liquid properties,  essential for circumventing
the explicit manybody calculations, are determined by using the
Luttinger sumrule as an input. Their phenomenological possibilities
could be demonstrated by leaving $Z$ as a free parameter.\cite{Matho,Matho1,Byczuk}
The cornerstone of the CFM as a
microscopic method is now the variation of the total energy to obtain $Z$.
Given $G(k,\omega)$, we calculate the total energy from the exact manybody
expression, actually another sumrule first found by Galitski.
However, without an explicit wavefunction, we have no rigorous variational
principle. In making the Gutzwiller approximation, beyond the
Gutzwiller ansatz for the wavefunction, one is also abandoning the
rigorous variational principle but one keeps $Z$ as variational
parameter.

The answer to the question, why our method  is variational, is
probably that we are using a GF, fully constrained by sumrules, that
leaves no other free parameter but $Z$. To obtain the Kondo effect,
we need degeneracy. Our model has spin degeneracy, $N_{f}=2$, for
its flavors. A clue, why including the incoherent background
spectrum, instead of projecting it out, improves the outcome for $Z$
comes from the limit of large $N_{f}$. \cite{Hewson,Rasul} The low
energy scale (Kondo temperature) in the Anderson impurity model can
be obtained exactly, as function of $n$ and $m=n/N_{f}$. Also,
coherent spectral weight is of order zero in $1/N_{f}$, the leading
background contribution starts at first order. Neglecting
background, as for instance in the slave boson method at mean-field
level, yields $Z=(1-n)/(1-m)$, as plotted for $N_{f}=2$ in Fig.\
\ref{fig3}. Compared with the Bethe ansatz, this renormalization is
not enough. Now, the influence of the background is strikingly
illustrated by solving for the Kondo temperature only to the first
order in $1/N_{f}$.\cite{Rasul} This causes indeed a substantial
decrease, bringing the result close to the exact value. The correct
doping dependence displays the upward curvature, as also seen in our
approximations CFM1 and CFM2. Finally, the improvement from CFM1 to
CFM2 shows a delicate interplay between the dynamical weight
transfer, related to the double occupancy, and the low energy scale.

If determining $Z$ by varying the total energy is indeed a valid
variational principle, it makes the CFM independent of the limit
$d=\infty$, thus giving it high flexibility and a large field of
applications. It is straightforward and, for low dimensional
systems, potentially very important to incorporate  the
$k-$dependence in the moment $M_{3}$. The  term $W_{3}(k)$ in
Eq.~(\ref{eq6a}) was already identified in the exact diagonalization
of a short linear chain, \cite{Mehlig} as causing a coupling of the
QP to antiferromagnetic fluctuations. This can be generalized to
fluctuations above other possible groundstates and the
selfconsistent determination of $W_{3}(k)$ thus offers a path to
describing the feedback of bosonic fluctuations on the low energy
sector. Up to now, the treatment of low energy effects within the
projection method was based more on physical intuition, or guesswork
for the more critical observer, than on an objective procedure.

The extension of the CFM to higher moments becomes possible due to its close
relationship with the numerical Lanczos procedure for finite
lattices. All what is missing is a proper termination of the
continued fraction with a TF representing the
low energy sector and the dissipation. The general algorithm for calculating the
coefficients in the TF is given in  Ref.\ \onlinecite{Matho}.

As an outlook, we enumerate other possibilities that are inherent in
the CFM, beyond the results of this paper. They are listed roughly
according to increasing effort that will be required to implement
them. (i) A more detailed exploitation of spectral functions on the
$k$-resolved level: e.g. the interpretation of Raman, ARPES, or
tunneling data requires the partial summation of $A(k,\epsilon)$
over selected spots in the Brillouin zone, weighted by matrix
elements. (ii) Hubbard lattice models with a more realistic kinetic
energy part, including van Hove singularities. (iii) The generalized
periodic Anderson model (PAM): lattice models with Hubbard repulsion
among transition orbitals but, in addition, hybridization with
ligand orbitals. (iv) Implementation of LDA+CFM. The algebraic
simplicity of the CFM allows to calculate the charge transfer
effects, present in model Hamiltonians of the PAM type, on an
"ab-initio" level. These effects, important for many real materials,
could not yet be handled successfully  by LDA+DMFT. (v) Not
difficult to implement, but leaving the strict framework of the CFM
as an algebraic method, is the inclusion of non Fermi liquid effects
on a phenomenological level. \cite{Matho}

In conclusion, we have attempted to demonstrate by means of the
Hubbard model that the CFM is a powerful method. Numerically simple, due to
its algebraic structure, it is still sufficiently rigorous to deal with strongly
correlated electrons in mesoscopic and macroscopic samples of condensed matter.

\end{document}